\documentclass[aps,prb,superscriptaddress,amsfonts,amsmath,amssymb,floatfix,nofootinbib,reprint,longbibliography]{revtex4-2}

\usepackage{booktabs}
\usepackage[dvipsnames]{xcolor}
\usepackage[colorlinks=true, urlcolor=blue, linkcolor=blue, citecolor=teal, pdftex]{hyperref}
\usepackage{mathtools}
\usepackage{siunitx}
\usepackage{tikz}

\usepackage{braket}

\usepackage{pgfplots}
\usepackage{pgfplotstable}
\pgfplotsset{width=\columnwidth,compat=1.9}

\usepgfplotslibrary{external}

\definecolor{googleB}{HTML}{4285F4}
\definecolor{googleG}{HTML}{34A853}
\definecolor{googleY}{HTML}{FBBC05}
\definecolor{googleR}{HTML}{EA4335}
\definecolor{googleBG}{HTML}{3B96A4}

\definecolor{mathmColor1}{rgb}{0.368417, 0.506779, 0.709798}
\definecolor{mathmColor2}{rgb}{0.880722, 0.611041, 0.142051}
\definecolor{mathmColor3}{rgb}{0.560181, 0.691569, 0.194885}
\definecolor{mathmColor4}{rgb}{0.922526, 0.385626, 0.209179}
\definecolor{mathmColor5}{rgb}{0.528488, 0.470624, 0.701351}
\definecolor{mathmColor6}{rgb}{0.772079, 0.431554, 0.102387}
\definecolor{mathmColor7}{rgb}{0.363898, 0.618501, 0.782349}
\definecolor{mathmColor8}{rgb}{1, 0.75, 0}
\definecolor{mathmColor9}{rgb}{0.647624, 0.37816, 0.614037}
\definecolor{mathmColor10}{rgb}{0.571589, 0.586483, 0.}
\definecolor{mathmColor11}{rgb}{0.915, 0.3325, 0.2125}
\definecolor{mathmColor12}{rgb}{0.400822, 0.522007, 0.85}
\definecolor{mathmColor13}{rgb}{0.972829, 0.621644, 0.073362}
\definecolor{mathmColor14}{rgb}{0.736783, 0.358, 0.503027}
\definecolor{mathmColor15}{rgb}{0.280264, 0.715, 0.429209}

\begin{document}

\title{Variational optimization of projected entangled-pair states on the triangular lattice}

\newcommand{\FUB}{Freie Universität Berlin, Dahlem Center for Complex Quantum Systems and Institut f\"{u}r Theoretische Physik, Arnimallee 14, 14195 Berlin, Germany}

\newcommand{\HZB}{Helmholtz-Zentrum Berlin für Materialien und Energie, Hahn-Meitner-Platz 1, 14109 Berlin, Germany}

\author{Jan Naumann}
\affiliation{\FUB}

\author{Jens Eisert}
\affiliation{\FUB}
\affiliation{\HZB}

\author{Philipp Schmoll}
\affiliation{\FUB}

\date{\today}

\makeatletter

\def\pgfplotstable@linear@regression#1{%
    \begingroup
    \pgfqkeys{/pgfplots/table/create col/linear regression}{/pgf/fpu,#1}%
    \pgfkeysgetvalue{/pgfplots/table/create col/linear regression/x}{\pgfplotstable@xsrc}%
    \pgfkeysgetvalue{/pgfplots/table/create col/linear regression/y}{\pgfplotstable@ysrc}%
    \pgfkeysgetvalue{/pgfplots/table/create col/linear regression/table}{\pgfplotstable@table}%
    \pgfkeysgetvalue{/pgfplots/table/create col/linear regression/xmode}{\pgfplotstable@xmode}%
    \pgfkeysgetvalue{/pgfplots/table/create col/linear regression/ymode}{\pgfplotstable@ymode}%
    \pgfkeysgetvalue{/pgfplots/table/create col/linear regression/variance}{\pgfplotstable@variance@colname}%
    \pgfkeysgetvalue{/pgfplots/table/create col/linear regression/variance list}{\pgfplotstable@variance@list}%
    \pgfkeysgetvalue{/pgfplots/table/create col/linear regression/variance src}{\pgfplotstable@variance@table}%
    \ifx\pgfplotstable@table\pgfutil@empty
        \pgfutil@ifundefined{pgfplotstablename}{}{
            \let\pgfplotstable@table=\pgfplotstablename
        }%
    \fi
    \ifx\pgfplotstable@table\pgfutil@empty
        \pgfplots@error{Sorry, I couldn't determine a value for create col/linear regression/table. Which table should I load?}%
    \fi
    \ifx\pgfplotstable@xsrc\pgfutil@empty
        \pgfplotsifinaddplottablestruct{%
            \pgfutil@ifundefined{pgfplots@plot@tbl@x}{}{%
                \let\pgfplotstable@xsrc=\pgfplots@plot@tbl@x
                \ifx\pgfplotstable@ysrc\pgfutil@empty
                    \pgfplotstablegetcolsof\pgfplots@table
                    \ifnum\pgfplotsretval=2
                    \else
                        \pgfplotsthrow{invalid argument}{\pgfplotstable@ysrc}{Sorry, I don't which column should be used as `y' for the linear regression. Please provide 'linear regression={y=<colname>}'}\pgfeov%
                    \fi
                \fi
            }%
        }{}%
    \fi
    \ifx\pgfplotstable@xsrc\pgfutil@empty
        \def\pgfplotstable@xsrc{[index]0}%
    \fi
    \ifx\pgfplotstable@ysrc\pgfutil@empty
        \def\pgfplotstable@ysrc{[index]1}%
    \fi
    \t@pgfplots@toka=\expandafter{\pgfplotstable@table}%
    \t@pgfplots@tokb=\expandafter{\pgfplotstable@xsrc}%
    \t@pgfplots@tokc=\expandafter{\pgfplotstable@ysrc}%
    \edef\pgfplots@loc@TMPa{{\the\t@pgfplots@tokb}\noexpand\of{\the\t@pgfplots@toka}}%
    \edef\pgfplots@loc@TMPb{{\the\t@pgfplots@tokc}\noexpand\of{\the\t@pgfplots@toka}}%
    \expandafter\pgfplotstablegetcolumn\pgfplots@loc@TMPa\to\pgfplotstable@X
    \expandafter\pgfplotstablegetcolumn\pgfplots@loc@TMPb\to\pgfplotstable@Y
    \edef\pgfplotstable@xmode{\pgfplotstable@xmode}%
    \expandafter\pgfplotstable@linear@regression@prepare@mode\expandafter{\pgfplotstable@xmode}{x}
    \edef\pgfplotstable@ymode{\pgfplotstable@ymode}%
    \expandafter\pgfplotstable@linear@regression@prepare@mode\expandafter{\pgfplotstable@ymode}{y}
    \ifx\pgfplotstable@variance@list\pgfutil@empty
        \pgfplotslistnewempty\pgfplotstable@VARIANCE
        \ifx\pgfplotstable@variance@colname\pgfutil@empty
        \else
            \ifx\pgfplotstable@variance@table\pgfutil@empty
                \t@pgfplots@toka=\expandafter{\pgfplotstable@table}%
                \t@pgfplots@tokb=\expandafter{\pgfplotstable@variance@colname}%
                \edef\pgfplots@loc@TMPa{{\the\t@pgfplots@tokb}\noexpand\of{\the\t@pgfplots@toka}}%
                \expandafter\pgfplotstablegetcolumn\pgfplots@loc@TMPa\to\pgfplotstable@VARIANCE
            \else
                \t@pgfplots@toka=\expandafter{\pgfplotstable@variance@colname}%
                \t@pgfplots@tokb=\expandafter{\pgfplotstable@variance@table}%
                \edef\pgfplotstable@loc@TMPa{%
                    \noexpand\pgfplotstablegetcolumn{\the\t@pgfplots@toka}\noexpand\of{\the\t@pgfplots@tokb}\noexpand\to\noexpand\pgfplotstable@VARIANCE}%
                \pgfplotstable@loc@TMPa
            \fi
        \fi
    \else
        \expandafter\pgfplotslistnew\expandafter\pgfplotstable@VARIANCE\expandafter{\pgfplotstable@variance@list}%
    \fi
    \pgfplotslistnewempty\pgfplotstable@Xparsed
    \pgfmathfloatcreate{0}{0.0}{0}%
    \let\pgfplotstable@S=\pgfmathresult
    \let\pgfplotstable@Sxx=\pgfmathresult
    \let\pgfplotstable@Sx=\pgfmathresult
    \let\pgfplotstable@Sy=\pgfmathresult
    \let\pgfplotstable@Sxy=\pgfmathresult
    \pgfutil@loop
    \pgfplotslistcheckempty\pgfplotstable@X
    \ifpgfplotslistempty
        \pgfplots@loop@CONTINUEfalse
    \else
        \pgfplots@loop@CONTINUEtrue
    \fi
    \ifpgfplots@loop@CONTINUE
        \pgfplotslistpopfront\pgfplotstable@X\to\pgfplotstable@x
        \pgfplotslistpopfront\pgfplotstable@Y\to\pgfplotstable@y
        \pgfplotstableparsex{\pgfplotstable@x}%
        \let\pgfplotstable@x=\pgfmathresult
        \expandafter\pgfplotslistpushback\pgfmathresult\to\pgfplotstable@Xparsed
        \pgfplotstableparsey{\pgfplotstable@y}%
        \let\pgfplotstable@y=\pgfmathresult
        \pgfmathfloatifflags{\pgfplotstable@y}{3}{}{
        \pgfplotslistcheckempty\pgfplotstable@VARIANCE
        \ifpgfplotslistempty
            \pgfmathfloatcreate{1}{1.0}{0}%
            \let\pgfplotstable@invsqr=\pgfmathresult
        \else
            \pgfplotslistpopfront\pgfplotstable@VARIANCE\to\pgfplotstable@variance
            \pgfmathfloatparsenumber{\pgfplotstable@variance}%
            \let\pgfplotstable@variance=\pgfmathresult
            \pgfmathfloatmultiply@{\pgfplotstable@variance}{\pgfplotstable@variance}%
            \let\pgfplotstable@sqr=\pgfmathresult
            \pgfmathfloatreciprocal@{\pgfplotstable@sqr}%
            \let\pgfplotstable@invsqr=\pgfmathresult
        \fi
        \pgfmathfloatadd@{\pgfplotstable@S}{\pgfplotstable@invsqr}%
        \let\pgfplotstable@S=\pgfmathresult
        \pgfmathfloatmultiply@{\pgfplotstable@x}{\pgfplotstable@invsqr}%
        \let\pgfplots@table@accum=\pgfmathresult
        \pgfmathfloatadd@{\pgfplotstable@Sx}{\pgfplots@table@accum}%
        \let\pgfplotstable@Sx=\pgfmathresult
        \pgfmathfloatmultiply@{\pgfplotstable@x}{\pgfplots@table@accum}%
        \let\pgfplots@table@accum=\pgfmathresult
        \pgfmathfloatadd@{\pgfplotstable@Sxx}{\pgfplots@table@accum}%
        \let\pgfplotstable@Sxx=\pgfmathresult
        \pgfmathfloatmultiply@{\pgfplotstable@y}{\pgfplotstable@invsqr}%
        \let\pgfplots@table@accum=\pgfmathresult
        \pgfmathfloatadd@{\pgfplotstable@Sy}{\pgfplots@table@accum}%
        \let\pgfplotstable@Sy=\pgfmathresult
        \pgfmathfloatmultiply@{\pgfplotstable@x}{\pgfplots@table@accum}%
        \let\pgfplots@table@accum=\pgfmathresult
        \pgfmathfloatadd@{\pgfplotstable@Sxy}{\pgfplots@table@accum}%
        \let\pgfplotstable@Sxy=\pgfmathresult
        }
    \pgfutil@repeat
    \pgfmathparse{\pgfplotstable@S * \pgfplotstable@Sxx - \pgfplotstable@Sx *\pgfplotstable@Sx}%
    \let\pgfplotstable@delta=\pgfmathresult
    \pgfmathparse{(\pgfplotstable@S * \pgfplotstable@Sxy - \pgfplotstable@Sx * \pgfplotstable@Sy) / \pgfplotstable@delta}%
    \let\pgfplotstable@a=\pgfmathresult
    \pgfmathparse{(\pgfplotstable@Sxx * \pgfplotstable@Sy - \pgfplotstable@Sx * \pgfplotstable@Sxy) / \pgfplotstable@delta}%
    \let\pgfplotstable@b=\pgfmathresult
    \pgfplotslistnewempty\pgfplotstable@RESULT
    \pgfplotslistforeachungrouped\pgfplotstable@Xparsed\as\pgfplotstable@x{%
        \pgfmathfloatmultiply@{\pgfplotstable@x}{\pgfplotstable@a}%
        \let\pgfplotstable@tmp=\pgfmathresult
        \pgfmathfloatadd@{\pgfplotstable@tmp}{\pgfplotstable@b}%
        \ifx\pgfplotstableparseylogbase\pgfutil@empty
        \else
            \pgfplotstableparseyinv@{\pgfmathresult}%
        \fi
        \pgfmathfloattosci{\pgfmathresult}%
        \expandafter\pgfplotslistpushback\pgfmathresult\to\pgfplotstable@RESULT
    }%
    \pgfmathfloattosci\pgfplotstable@a
    \let\pgfplotstable@a=\pgfmathresult
    \pgfmathfloattosci\pgfplotstable@b
    \let\pgfplotstable@b=\pgfmathresult
    \global\let\pgfplotstableregressiona\pgfplotstable@a%
    \global\let\pgfplotstableregressionb\pgfplotstable@b%
    \let\pgfplotsretval=\pgfplotstable@RESULT
    \pgfmath@smuggleone\pgfplotsretval
    \endgroup
}%

\pgfplotstableread[col sep=&,]{data/data_TL.txt}\datatableTL%

\pgfplotstableread[col sep=&,]{data/data_KL.txt}\datatableKL%

\pgfplotsset{select coords between index/.style 2 args={
    x filter/.code={
        \ifnum\coordindex<#1\def\pgfmathresult{}\fi
        \ifnum\coordindex>#2\def\pgfmathresult{}\fi
    }
}}

\makeatother

\begin{abstract}

We introduce a general corner transfer matrix renormalization group algorithm tailored to projected entangled-pair states on the triangular lattice.
By integrating automatic differentiation, our approach enables direct variational energy minimization on this lattice geometry.
In contrast to conventional approaches that map the triangular lattice onto a square lattice with diagonal next-nearest-neighbour interactions, our native formulation yields improved variational results at the same bond dimension. This improvement stems from a more faithful and physically informed representation of the entanglement structure in the tensor network and an increased number of variational parameters.
We apply our method to the antiferromagnetic nearest-neighbour Heisenberg model on the triangular and kagome lattice, and benchmark our results against previous numerical studies.

\end{abstract}

\maketitle

\section{Introduction}

The study of condensed matter and quantum many-body physics critically depends on the development of efficient classical simulation techniques to address open questions and to uncover novel phenomena. 
In this pursuit, \emph{tensor network} (TN) methods~\cite{Orus2014,Orus2019,Cirac2021,Eisert2010,Bridgeman2017} have emerged as an indispensable set of techniques, rooted in the principles of quantum information theory and distinguished by their ability to faithfully represent the entanglement structure inherent in complex quantum systems. 
Initially introduced through the one-dimensional \emph{density matrix renormalization group} (DMRG)~\cite{White92} and its formulation in terms of \emph{matrix product states} (MPS)~\cite{Oestlund1995,Dukelsky1998,PerezGarcia2007,Schollwoeck2011}, tensor networks have since evolved into powerful tools applicable in higher spatial dimensions. 
In particular, \emph{projected entangled-pair states} (PEPS)~\cite{Verstraete2004} have established themselves 
as one of the most important frameworks for simulating two-dimensional quantum systems, with significant progress made by incorporating direct variational energy optimization~\cite{Corboz2016,Vanderstraeten2016,Rader2018}. 
A key development in this context has been the use of \emph{automatic differentiation} (AD)~\cite{Liao219,Naumann2024,Francuz2025} to efficiently and accurately compute energy gradients, significantly simplifying the optimization process. 

These methodological improvements have notably enhanced the applicability and precision of PEPS in challenging settings such as frustrated quantum magnetism~\cite{Hasik2021,Hasik2022,Niu2022,Schmoll2023,Francesco2023,Zhang2023,Xu2023,Weerda2024,Schmoll2024,corboz2025quantumspinliquidphase}, where subtle competition between exotic phases can now be resolved with greater confidence. 
Naturally, PEPS has been applied to the study of paradigmatic models, such as the spin-$1/2$ Heisenberg model on the triangular~\cite{Li2022,Chi2022,Hasik2024} and kagome~\cite{Xie2014,Picot2016,Liao2017,Jiang2019,Schmoll2020_3} lattices. 
In both systems, quantum fluctuations are significantly enhanced due to strong geometric frustration on the triangular motifs. 
This, in turn, typically leads to the aforementioned subtle competition of candidate states that is to be resolved on small energy scales. 
While the ground state of the triangular lattice Heisenberg model is known to have $120^\circ$ in-plane magnetic order~\cite{Capriotti1999,White2007}, its counterpart on the kagome lattice does not magnetically order down to zero temperature and is conjectured to realize a \emph{quantum spin liquid} (QSL) state~\cite{Yan2011,Depenbrock2012,Iqbal2013,Iqbal2014,Liao2017,Jiang2019,Laeuchli2019}. 

To date, the variational optimization of projected entangled-pair states on lattices with an underlying triangular Bravais lattice has almost exclusively been performed on a square lattice geometry, due to its simplicity for all involved algorithmic steps, in particular the \emph{corner transfer matrix renormalization group} (CTMRG) algorithm~\cite{Nishino1996,Nishino1997,Orus2009,Corboz2014,Fishman2018}. 
To this end, the triangular lattice is treated as a square lattice with next-nearest neighbour interactions, so that correlations along this diagonal lattice direction need to be routed along the remaining ones. 
Here, instead, we introduce a computationally efficient and accurate method for contracting and variationally optimizing projected entangled-pair states directly on the triangular lattice. 
This development comes with significant advantages in terms of representability of the underlying entanglement structure due to the increased number of links in the ansatz, as well as a larger number of variational parameters at the same PEPS bond dimension. 
On the downside, it also leads to an unavoidable increase in computational complexity compared to the conventional, square lattice CTMRG. 
We benchmark our algorithm on the antiferromagnetic spin-$1/2$ Heisenberg model on the triangular and kagome lattices, with the latter involving only a mild additional coarse-graining overhead.
Our results are among the lowest variational energies reported for those models, and support the general picture from previous studies in the literature. 
Even with rather low accessible bond dimensions, the findings demonstrate the enormous potential of our methods for addressing challenging problems on triangular-based structures.

\section{Triangular Lattice CTMRG}
\label{sec:triangularLatticeCTMRG}

Building on the proposal in Ref.~\cite{Lukin2024} but going substantially beyond this work, we define a corner transfer matrix renormalization group method tailored to the triangular lattice. Giving up on rotation and reflection symmetry of the input tensors, our procedure can accommodate arbitrary triangular projected entangled-pair states, including symmetry-broken and spiral states~\cite{Hasik2024}. 
Additionally, the algorithm can be conveniently extended to incorporate physical symmetries, which improves computational efficiency~\cite{Singh2010,Weichselbaum2012,Silvi2019,Schmoll2020_1,Mortier2025}. 
Conceptually, the overall scheme follows the well-established CTMRG on the square lattice. 
However, the triangular lattice variant comprises some differences, and has an inherently higher computational complexity to it.

\subsection{Triangular lattice PEPS}
\label{sub:triangularLatticePEPS}

Translationally invariant tensor network states capture quantum ground states of locally interacting Hamiltonians well. Specifically, infinite projected entangled-pair states represent state vectors of quantum many-body lattice systems
\begin{align}
    \ket{\psi} = \sum_{\lbrace s_{\mathbf r} \rbrace} C_{\lbrace s_{\mathbf r} \rbrace} \ket{\lbrace s_{\mathbf r} \rbrace}
    \label{eq:manyBodyStateVector}
\end{align}
in the thermodynamic limit, where the coefficients $C_{\lbrace s_{\mathbf r} \rbrace}$ are encoded in the contraction of an (infinite) two-dimensional tensor network. 
The archetypal projected entangled-pair state on the \emph{square 
lattice} (SL) is trivially generalized to the \emph{triangular lattice} (TL), where it is generated by a set of seven-index tensors that are periodically repeated according 
to
\begin{align}
    \begin{split}
        \includegraphics[scale = 1.0]{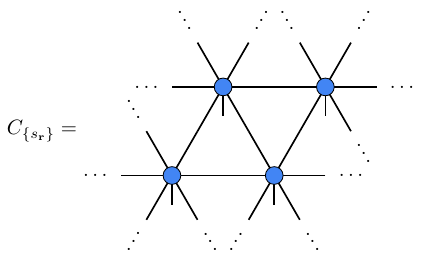}
    \end{split}.
    \label{eq:triangularPEPS_1}
\end{align} 
The open vertical legs of dimension $p$ represent the physical degrees of freedom $\lbrace s_{\mathbf r} \rbrace$ of the coefficient tensor in Eq.~\eqref{eq:manyBodyStateVector}, while the virtual legs connect to all neighbouring sites in the triangular lattice. 
They are used to mediate quantum correlations in the system between particles at different positions $\mathbf r = (x, y)$. 
The bond dimension of the virtual bulk indices, denoted by $\chi_B$, is the main refinement parameter in the PEPS ansatz that controls the maximal amount of entanglement. 
Each tensor holds $p\chi_B^6$ variational parameters, which are optimized so that the infinite PEPS is the best approximation to the target state of interest at the specific bond dimension.\\

The TL ansatz exhibits profound physical and computational differences compared to the established SL PEPS. 
From both a physical and computational perspective, it is much more efficient at capturing the quantum correlations along the three independent lattice directions, which makes it particularly well-suited for numerical applications. 
Moreover, by incorporating the actual geometry of the triangular lattice, it is expected to represent ground states more naturally, especially those with corresponding rotational symmetry. 
While both the PEPS ansatz on the SL and TL satisfy the area law for the entanglement entropy by construction~\cite{Orus2014}, the entanglement entropy for a bipartition between a patch of $L \times L$ tensors and its environment is significantly different. 
Specifically, it is upper bounded by
\begin{align}
    S(L) = 
    \begin{cases}
        \hfil 4L \log \chi_B & \text{for the SL},\\
        (8L - 2) \log \chi_B & \text{for the TL}.
    \end{cases}
\end{align}
In the limit of large $L$, we find that the upper bound for the TL is twice as large as for the SL, so that a SL PEPS requires the square of the bond dimension of a TL PEPS to reproduce it. 
This has direct physical consequences for TL PEPS in terms of representability and accuracy, as we will demonstrate below. 
These bounds can be obtained by counting how many virtual indices are cut when separating the $L \times L$ patch from the full tensor network.

\begin{figure}[ht]
    \centering
    \tikzsetnextfilename{comparisonVariationalParameters}
\begin{tikzpicture}
        
    \begin{axis}[
        xlabel=$\chi_{B}$, 
        ylabel=$N_\text{var}$, 
        width = 1.00\columnwidth,
        height = 1.00\columnwidth,
        label style={font=\small},
        tick label style={font=\footnotesize},
        every axis/.append style={thick}, 
        xtick = {2, 3, 4, 5, 6, 7, 8}, 
        xticklabels = {$2$, $3$, $4$, $5$, $6$, $7$, $8$}, 
        ymode = log,
        log ticks with fixed point,
        ymin = 1e0, 
        ymax = 1e6, 
        ytick = {1e0, 1e1, 1e2, 1e3, 1e4, 1e5, 1e6}, 
        yticklabels = {$10^{0}$, $10^{1}$, $10^{2}$, $10^{3}$, $10^{4}$, $10^{5}$, $10^{6}$},
        legend style = {at = {(0.05, 0.85)}, anchor = west}, 
        width=\columnwidth,
        height=0.75\columnwidth,
    ]

    \addplot[domain = 2:8, samples=50, smooth, googleB, line width = 1.5] {2*x^4};
    \addlegendentry{SL, $p\chi_{B}^4$}
    
    \addplot[domain = 2:8, samples=50, smooth, googleG, line width = 1.5] {2*x^6};
    \addlegendentry{TL, $p\chi_{B}^6$}

    
    \addplot[mark=none, gray, dotted] coordinates {(2.0, 128) ({2*sqrt(2)}, 128)};
    \addplot[mark=none, gray, dotted] coordinates {({2*sqrt(2)}, 1) ({2*sqrt(2)}, 128)};

    \addplot[mark=none, gray, dotted] coordinates {(3.0, 1458) (5.196, 1458)};
    \addplot[mark=none, gray, dotted] coordinates {(5.196, 1) (5.196, 1458)};

    \addplot[mark=none, gray, dotted] coordinates {(4.0, 8192) (8.0, 8192)};
    \addplot[mark=none, gray, dotted] coordinates {(8, 1) (8, 8192)};
    
    \end{axis}

\end{tikzpicture}
    \caption{Comparison of the number of variational parameters in a single real tensor for projected entangled-pair states defined on the square lattice (SL) and triangular lattice (TL), assuming a physical dimension of $p = 2$.}
    \label{fig:comparisonVariationalParameters}
\end{figure}
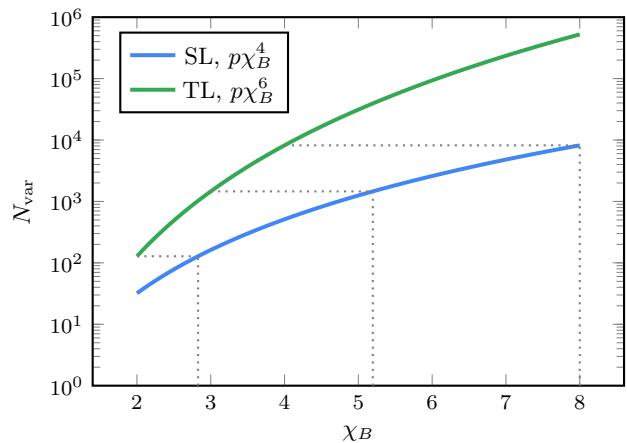

From a computational perspective however, the number of tensor indices and their dimensions play a pivotal role.
In Fig.~\ref{fig:comparisonVariationalParameters} we show a  comparison for the number of variational parameters of a single real PEPS tensor for both variants, fixing $p = 2$ as the physical dimension of a spin-$1/2$.
While fourth- and sixth-order scalings are clearly drastically different, it is instructive to note that a TL tensor at \mbox{$\chi_B = 4$} holds as many variational parameters as a SL tensor at \mbox{$\chi_B = 8$}. 
This larger number of parameters at the same bond dimension can in principle be advantageous for variational optimization. 
However, the increased number of tensor legs in all algorithmic steps increases the computational complexity, which in turn limits the maximal accessible bond dimensions compared to the SL case.

One of the central objects in the numerical simulation of quantum many-body lattice systems using projected entangled-pair states is the double-layer tensor network of the norm, i.e.,
\begin{align}
    \braket{\psi \vert \psi} = \operatorname{tTr}\left( \prod_{\mathbf r} A_{\mathbf r} \cdot A_{\mathbf r}^\star \right).
\end{align}
Here, $\operatorname{tTr}$ denotes the tensor trace over all virtual indices, and each local PEPS tensor $A_{\mathbf r}$ is contracted with its complex conjugate over the physical index $s_{\mathbf r}$. 
This contraction can not be performed exactly without an exponential increase in bond dimension and 
comes along with rigorous obstructions of computational complexity in worst and average case~\cite{Schuch2007,Haferkamp2020} (even though for gapped models), so that approximate methods have to be used in practice. 
Here, we make use of the corner transfer matrix renormalization group scheme, one of the long-established techniques to contract infinite PEPS. 
The CTMRG is an iterative power method, which will compute effective fixed-point environment tensors that contain semi-infinite parts of the contracted network.

\subsection{Environment tensor definition}
\label{sub:environmentTensorDefinition}

\begin{figure}[b!]
    \centering
    \includegraphics[scale = 1.0]{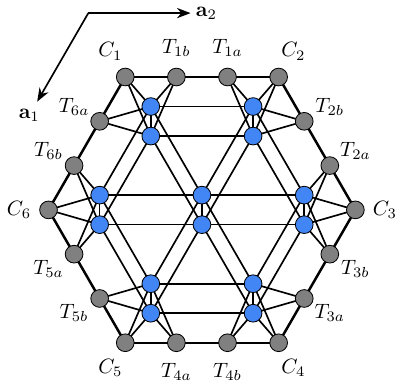}
    \caption{Definition of all six corner tensors $C$ and twelve edge tensors $T$ in the triangular CTMRG procedure. The triangular lattice is spanned by lattice vectors $\mathbf a_1$ and $\mathbf a_2$. The edge tensors are labeled such that, in each of the six lattice directions, both tensors connect to the same PEPS tensor and $T_{i,a}$ appears before $T_{i,b}$ when traversing the network clockwise.}
    \label{fig:tensorDefinition_CTMRG_0}
\end{figure}

\begin{figure}[ht]
    \centering
    \includegraphics[scale = 1.0]{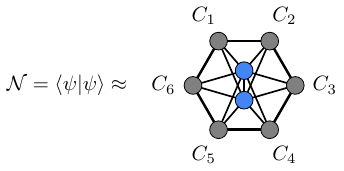}
    \caption{Single-site norm of the PEPS state vector computed from the minimal set of CTMRG environment tensors, which only includes all six corner tensors.}
    \label{fig:tensorDefinition_CTMRG_1}
\end{figure}

In order to define the \emph{radial} CTMRG procedure on the triangular lattice, we assume a fully translational invariant infinite PEPS ansatz. 
Since our method does not assume rotational symmetry of the input PEPS tensors, the CTMRG environment consists of six corner tensors $C$, as well as twelve edge tensors $T$. 
The full definition of those boundary tensors is shown in Fig.~\ref{fig:tensorDefinition_CTMRG_0}. 
Here, the perimeter bond dimension $\chi_E$ controls the amount of approximations in the triangular CTMRG. 
In practice, it needs to be chosen sufficiently high for expectation values to be converged. 
Note, that \emph{all} environment tensors are rank-four tensors, unlike in the square lattice CTMRG where corner tensors are only matrices. 
The environment tensors generated in the iterative CTMRG power method are primarily used to compute the norm of the state vector, as well as expectation value of local observables. 
However, they can also be used to extract other quantities, such as the effective correlation length $\xi(\chi_E)$ introduced by approximating the contraction of the infinite PEPS network. 
The norm of the PEPS for a single-site observable can be computed from just the corner environments, as shown in Fig.~\ref{fig:tensorDefinition_CTMRG_1}.
Nearest-neighbour expectation values on the triangular lattice can then be computed from a two-site network in all three lattice directions, i.e.,  $\mathbf a_1$, $\mathbf a_2$ and $\mathbf a_1 + \mathbf a_2$. 
Operators with a larger support can be computed similarly, by considering the corresponding larger patch of the infinite PEPS network including the surrounding environment tensors.

The triangular lattice CTMRG algorithm can be similarly implemented for non-trivial unit cells. 
In this case, all local PEPS and environment tensors have a position label $(x, y)$ assigned to them, and all algorithmic steps need to respect the lattice periodicity. 
The complete triangular CTMRG algorithm, including the absorption steps and calculation of different projectors is explained in full detail in the Appendix. 
The dominant computational cost of the triangular CTMRG is $\mathcal{O}(\chi_E^2 \chi_B^9 p + \chi_E^3 \chi_B^6)$, which, for a small physical dimension $p < \chi_B$ and environment bond dimensions $\chi_E \sim \chi_B^2$ scales as $\mathcal{O}(\chi_B^{13}p)$. 
This high scaling, combined with an observed necessity for even larger environment bond dimensions, unfortunately limits the accessible PEPS bond dimensions compared to a square lattice CTMRG.

\subsection{Comparison to square lattice PEPS}

To illustrate the practical impact of the different computational scalings, we provide a comparison of representative wall-clock runtimes. 
For a triangular lattice PEPS ansatz with physical dimension $p = 2$, bulk bond dimension $\chi_B = 4$ and environment dimension $\chi_E = 130$, the CTMRG algorithm introduced in this work needs approximately $119\,\text{s}$ per absorption step on a modern CPU. Under the same parameter choices, the CTMRG for a square lattice PEPS ansatz only requires about $17\,\text{s}$ per step and is therefore significantly faster. However, directly comparing triangular and square PEPS at the same bond dimension is misleading here. As discussed in the previous section, the TL PEPS has more variational parameters at fixed $\chi_B$ and, importantly, features enhanced entanglement scaling. A more meaningful comparison is therefore to match the number of local variational parameters. This is achieved by a SL PEPS with $\chi_B = 8$, which still has a slightly worse entanglement scaling,
\begin{align}
    \frac{S_\text{TL}(L, \chi_B=4)}{S_\text{SL}(L, \chi_B=8)} = \frac{4}{3} - \frac{1}{3L}.
\end{align}
In this setting, the square lattice CTMRG runtime increases to roughly $1525\,\text{s}$ per absorption step, which in turn is significantly slower compared to the TL benchmark.

Notice that while the choices above allow for a reasonably fair runtime comparison, the corresponding comparison of the actual quantum states is not obvious. Although one can formally map a TL tensor with bond dimension $\chi_B = 4$ to a SL tensor with $\chi_B = 8$ by reshaping its indices, this transformation does not preserve the entanglement structure of the state -- the resulting SL PEPS represents a quantum state with fundamentally different expressive power. Consequently, any benchmark based on such a mapping would intertwine changes in the underlying physical state with changes in the contraction scheme.

\section{Benchmarking results}
\label{sec:benchmarkingResults}

\begin{figure}[ht]
    \centering
    \includegraphics[width=0.65\columnwidth]{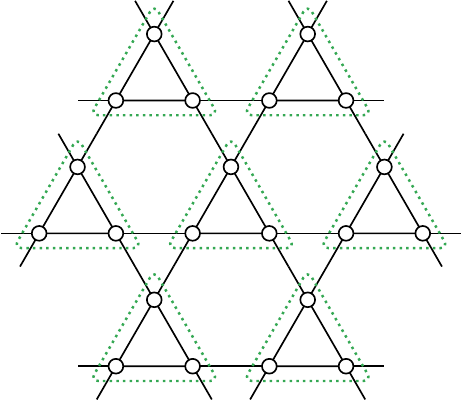}
    \caption{The kagome lattice has an underlying triangular Bravais lattice with a three-site basis, highlighted with green triangles. It can be simulated with a TL PEPS ansatz and a physical dimension of $p^3 = 8$ for a spin-$1/2$ system.}
    \label{fig:kagomeLattice_9}
\end{figure}

In order to benchmark the native variational optimization on the triangular lattice and compare it to the established square lattice version, we consider two paradigmatic antiferromagnetic spin models. 
The first one is the nearest-neighbour Heisenberg model on the triangular lattice, which is known to have a $120^\circ$ long-range magnetically ordered ground state~\cite{Capriotti1999,White2007}. 
The second one is the nearest-neighbour Heisenberg model on the kagome lattice, which has an underlying triangular Bravais structure with a three-site basis, see Fig.~\ref{fig:kagomeLattice_9}.
Its ground state is known to be a quantum spin liquid, for which competing methods suggest either a gapped $\mathbb{Z}_2$~\cite{Yan2011,Depenbrock2012,Laeuchli2019} or a gapless~\cite{Iqbal2013,Iqbal2014,Liao2017,Jiang2019} version. 

Both spin models are numerically challenging to simulate, as the high amount of frustration leads to close competition among many possible ground states, that has to be resolved at very small energy scales. 
A key requirement for the PEPS optimization on the triangular lattice is therefore the use of large environment bond dimensions. 
This is not only essential to ensure accurate expectation values but also for a reliable variational optimization, since too severe truncation of $\chi_E$ can lead to discontinuities in the approximate energy expectation value as a function of the PEPS parameters~\cite{Fedorovich2025}. 
In particular, the variational energy optimization is performed with a maximal environment bond dimension of $\chi_M = 150$ for all values of the PEPS bond dimension $\chi_B$. 
The actual bond dimension used in the optimization is dynamically increased with a truncation error threshold of $\epsilon = 10^{-7}$ up to the maximal value~\cite{Naumann2024}. 
Therefore, it can be smaller in practice if the truncation errors in the CTMRG projector calculation are below $\epsilon$, and thus insignificant. 
Unfortunately, the large environment bond dimensions reach the regime $\chi_E \sim \chi_B^3$, which adds to the already increased computational cost (cf.~Sec.~\ref{sub:environmentTensorDefinition}). 

We compare our native optimization on the TL against the SL variant in terms of ground state energy $E_0$, as well as the magnetic order parameter $m$. 
To introduce as little bias as possible, all bond dimensions are chosen to be the same in both the SL and TL simulations. 
For the data analysis, we then evaluate the expectation values at increasingly large environment bond dimensions, until convergence is reached. 
This eliminates bias from residual approximations in the CTMRG routine.

\subsection{Triangular lattice Heisenberg model}

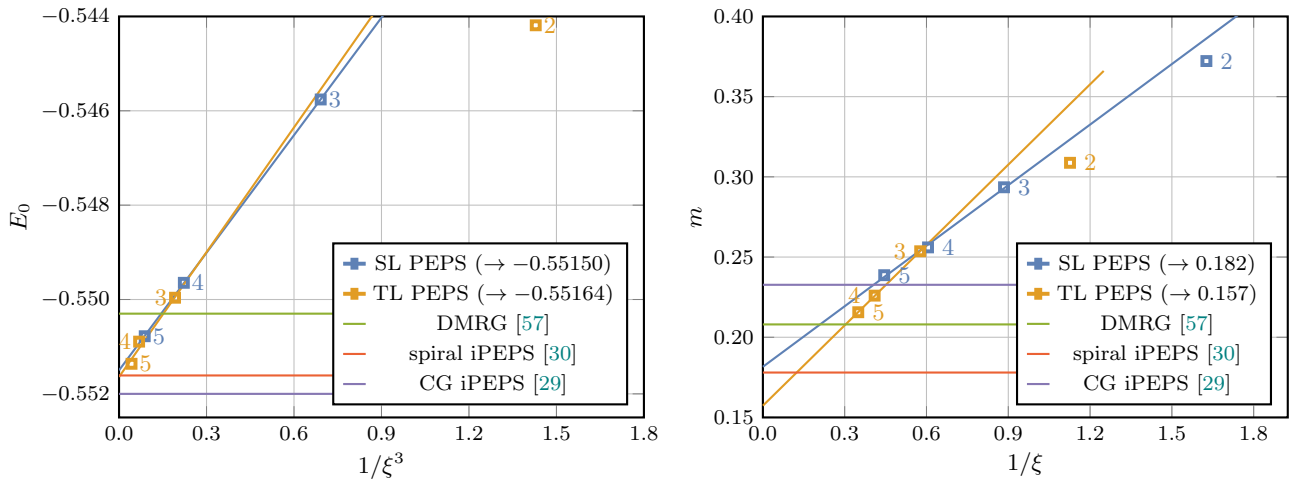
\begin{figure*}[ht]
    \centering
    \tikzsetnextfilename{results_TL}
\begin{tikzpicture}
    
    \begin{axis}[
        xlabel=$1/\xi^3$,
        xlabel shift=-2,
        ylabel=$E_0$,
        ylabel shift=-2,
        width=0.99\columnwidth,
        height=0.8\columnwidth,
        legend style={
            font=\footnotesize, 
        },
        label style={font=\small},
        tick label style={font=\footnotesize},
        every axis/.append style={thick},
        /tikz/mark size=1pt,
        legend pos=south east,
        xmajorgrids=true,
        ymajorgrids=true,
        xmin = 0.0, 
        xmax = 1.8, 
        xtick = {0.0, 0.3, 0.6, 0.9, 1.2, 1.5, 1.8}, 
        xticklabels = {$0.0$, $0.3$, $0.6$, $0.9$, $1.2$, $1.5$, $1.8$}, 
        ymin = -0.5525, 
        ymax = -0.544, 
        ytick = {-0.554, -0.552, -0.550, -0.548, -0.546, -0.544},
        yticklabels = {$-0.554$, $-0.552$, $-0.550$, $-0.548$, $-0.546$, $-0.544$}, 
    ]

    \addplot[
        color = mathmColor1, 
        dashed,
        mark = square, 
        mark options=solid,
        line width=1.5pt, 
        mark size=1.5pt, 
        only marks,
        point meta = explicit symbolic,
        visualization depends on = {value \thisrow{D} \as \mylabel},
        visualization depends on={\thisrow{angleSL} \as \myangle},
        every node near coord/.append style={%
            anchor=center, label={[label distance=-0.5em]\myangle:{\mylabel}}%
        },
        nodes near coords,
        legend image post style={sharp plot}
    ]
    table[
        x expr = 1/(\thisrow{xiSL}^3),
        y = ESL,
        col sep=&,
    ] {data/data_TL.txt};
    
    \addplot[
        draw = none,
        no markers,
        forget plot,
    ]
    table[
        x expr = 1/(\thisrow{xiSL}^3),
        y = {create col/linear regression = {y = ESL}},
        col sep=&,
    ] {data/data_TL_extrapolation.txt};
    \xdef\zeroTLSL{\pgfplotstableregressionb};

    \addplot[mathmColor1, domain=0.0 : 1.8, forget plot, dashed] {\pgfplotstableregressiona * x + \zeroTLSL};
    \addlegendentry{SL PEPS ($\rightarrow \pgfmathprintnumber[/pgf/number format/fixed,/pgf/number format/precision=5,/pgf/number format/fixed zerofill=true]{\zeroTLSL}$)};

    \addplot[
        color = mathmColor2,
        mark = triangle,
        mark options=solid,
        line width=1.5pt, 
        mark size=1.5pt, 
        only marks,
        point meta = explicit symbolic,
        visualization depends on = {value \thisrow{D} \as \mylabel},
        visualization depends on={\thisrow{angleTL} \as \myangle},
        every node near coord/.append style={%
            anchor=center, label={[label distance=-0.5em]\myangle:{\mylabel}}%
        },
        nodes near coords,
        legend image post style={sharp plot}
    ]
    table[
        x expr = 1/(\thisrow{xiTL}^3),
        y = ETL,
        col sep=&,
    ] {data/data_TL.txt};
    \addplot[
        draw = none,
        no markers,
        forget plot,
    ]
    table[
        col sep=&,%
        x expr = 1/(\thisrow{xiTL}^3),
        y = {create col/linear regression = {y = ETL}},
    ] {data/data_TL_extrapolation.txt};
    \xdef\zeroTLTL{\pgfplotstableregressionb};

    \addplot[mathmColor2, domain=0.0 : 1.8, forget plot] {\pgfplotstableregressiona * x + \zeroTLTL};
    \addlegendentry{TL PEPS ($\rightarrow \pgfmathprintnumber[/pgf/number format/fixed,/pgf/number format/precision=5,/pgf/number format/fixed zerofill=true]{\zeroTLTL}$)};



    \addplot[mark=none, mathmColor3] coordinates {(0, -0.5503) (1.6, -0.5503)};
    \addlegendentry{DMRG~\cite{Huang2024}};

    \addplot[mark=none, mathmColor4] coordinates {(0, -0.55161) (1.6, -0.55161)};
    \addlegendentry{spiral iPEPS~\cite{Hasik2024}};

    \addplot[mark=none, mathmColor5] coordinates {(0, -0.552) (1.6, -0.552)};
    \addlegendentry{CG iPEPS~\cite{Chi2022}};

    \end{axis}

    \begin{axis}[
        xlabel=$1/\xi$,
        xlabel shift=-2,
        ylabel=$m$,
        ylabel shift=-2,
        width=0.99\columnwidth,
        height=0.8\columnwidth,
        legend style={
            font=\footnotesize, 
        },
        label style={font=\small},
        tick label style={font=\footnotesize},
        every axis/.append style={thick},
        /tikz/mark size=1pt,
        legend pos=south east,
        xmajorgrids=true,
        ymajorgrids=true,
        xmin = 0.0, 
        xtick = {0.0, 0.3, 0.6, 0.9, 1.2, 1.5, 1.8}, 
        xticklabels = {$0.0$, $0.3$, $0.6$, $0.9$, $1.2$, $1.5$, $1.8$}, 
        ymin = 0.15, 
        ymax = 0.4, 
        ytick = {0.15, 0.20, 0.25, 0.30, 0.35, 0.40},
        yticklabels = {$0.15$, $0.20$, $0.25$, $0.30$, $0.35$, $0.40$},
        xshift=0.99\columnwidth,
    ]

    \addplot[
        color = mathmColor1, 
        dashed,
        mark = square, 
        mark options=solid,
        line width=1.5pt, 
        mark size=1.5pt,
        only marks,
        point meta = explicit symbolic,
        visualization depends on = {value \thisrow{D} \as \mylabel},
        visualization depends on={\thisrow{angleSL} \as \myangle},
        every node near coord/.append style={%
            anchor=center, label={[label distance=-0.25em]\myangle:{\mylabel}}%
        },
        nodes near coords,
        legend image post style={sharp plot}
    ]
    table[
        x expr = 1/(\thisrow{xiSL}),
        y = mSL,
        col sep=&
    ] {data/data_TL.txt};
    \addplot[
        color = mathmColor1, 
        mark = square, 
        line width=1.5pt, 
        mark size=1.5pt,
        draw=none,
        no marks,
        forget plot
    ]
    table[
        x expr = 1/(\thisrow{xiSL}),
        y = {create col/linear regression = {y = mSL}},
        col sep=&
    ] {data/data_TL_extrapolation.txt};
    \xdef\zeroMSL{\pgfplotstableregressionb};

    \addplot[mathmColor1, domain=0.0 : 1.75, forget plot, no marks, dashed] {\pgfplotstableregressiona * x + \zeroMSL};
    \addlegendentry{SL PEPS ($\rightarrow \pgfmathprintnumber[/pgf/number format/fixed,/pgf/number format/precision=3,/pgf/number format/fixed zerofill=true]{\zeroMSL}$)};

    \addplot[
        color = mathmColor2, 
        mark = triangle,
        mark options=solid,
        line width=1.5pt, 
        mark size=1.5pt,
        only marks,
        point meta = explicit symbolic,
        visualization depends on = {value \thisrow{D} \as \mylabel},
        visualization depends on={\thisrow{angleTL} \as \myangle},
        every node near coord/.append style={%
            anchor=center, label={[label distance=-0.25em]\myangle:{\mylabel}}%
        },
        nodes near coords,
        legend image post style={sharp plot}
    ]
    table[
        col sep=&,%
        x expr = 1/(\thisrow{xiTL}),%
        y = mTL,
    ] {data/data_TL.txt};
    \addplot[
        color = mathmColor2, 
        mark = square, 
        line width=1.5pt, 
        mark size=1.5pt,
        draw=none,
        no marks,
        forget plot
    ]
    table[
        col sep=&,%
        x expr = 1/(\thisrow{xiTL}),%
        y = {create col/linear regression = {y = mTL}},
    ] {data/data_TL_extrapolation.txt};
    \xdef\zeroMTL{\pgfplotstableregressionb};

    \addplot[mathmColor2, domain=0:1.25, forget plot, no marks] {\pgfplotstableregressiona * x + \zeroMTL};
    \addlegendentry{TL PEPS ($\rightarrow \pgfmathprintnumber[/pgf/number format/fixed,/pgf/number format/precision=3,/pgf/number format/fixed zerofill=true]{\zeroMTL}$)};



    \addplot[mark=none, mathmColor3] coordinates {(0, 0.208) (1.75, 0.208)};
    \addlegendentry{DMRG~\cite{Huang2024}};

    \addplot[mark=none, mathmColor4] coordinates {(0, 0.178) (1.75, 0.178)};
    \addlegendentry{spiral iPEPS~\cite{Hasik2024}};

    \addplot[mark=none, mathmColor5] coordinates {(0,  0.2327) (1.75,  0.2327)};
    \addlegendentry{CG iPEPS~\cite{Chi2022}};

    \end{axis}

\end{tikzpicture}
    \caption{Ground state energy $E_0$ and magnetic order parameter $m$ for the nearest-neighbour Heisenberg model on the triangular lattice. The data points are labeled by the PEPS bond dimension $\chi_B$ for a better comparison. Our results are compared against DMRG~\cite{Huang2024}, spiral iPEPS~\cite{Hasik2024} and coarse-grained iPEPS~\cite{Chi2022}.}
    \label{fig:results_TL}
\end{figure*}

The spin-$1/2$ antiferromagnetic Heisenberg model on the triangular lattice plays an important role in condensed matter physics and is a classic example featuring geometric frustration.  Its ground state is well-known to form long-range magnetic order, where the spins align in-plane with an angle of $120^\circ$ on three sublattices. 
The ordering wave vector \mbox{$\mathbf k = (2\pi/3, 2\pi/3)$} can be conveniently incorporated into the spiral PEPS ansatz~\cite{Hasik2024}, which naturally extends to the triangular lattice as well. 
Therefore, the model can be simulated with a single-site PEPS ansatz at fixed ordering wave vector. 

Due to gapless low-energy excitations~\cite{Zheng2006,Drescher2025} present, the environment bond dimension $\chi_E$ necessarily introduces a finite correlation length, whose effect can be eliminated by means of a \emph{finite correlation length scaling} (FCLS) analysis. 
To this end, the correlation length $\xi(\chi_E)$ is computed from the eigenvalue spectrum of the transfer matrix and extrapolated to the \mbox{$\chi_E \rightarrow \infty$} limit according to Ref.~\cite{Rader2018}. 

Table~\ref{tab:energyOrderParameter_TL} lists the ground state energy and magnetic order parameter obtained from variational optimization.
\begin{table}[ht]
    \centering
    \begin{filecontents*}{data/data_TL.txt}
D & xiSL & ESL & mSL & xiTL & ETL & mTL & angleSL & angleTL
2 & 0.6147733917191697 & -0.52675935 & 0.37219563 & 0.887826026389817 & -0.544187557 & 0.30875602243 & 0 & 0
3 & 1.130768085996412 & -0.54575992 & 0.29350136 & 1.7332667881843447 & -0.5499622616433772 & 0.25360859 & 0 & 180
4 & 1.650996809485430 & -0.54964623 & 2.56020429e-01 & 2.4396612516583334 & -0.55089267 & 2.25946210e-01 & 0 & 180
5 & 2.2473203196307567 & -0.55077832 & 2.38691718e-01 & 2.8588035835758023 & -0.5513626 & 2.15605156e-01 & 0 & 0
\end{filecontents*}
\pgfplotstabletypeset[%
        col sep=&,%
        columns={D, xiSL, ESL, mSL, xiTL, ETL, mTL},%
        /pgf/number format/fixed,%
        /pgf/number format/precision=5,%
        /pgf/number format/fixed zerofill=true,%
        every head row/.style={after row=\hline},%
        columns/D/.style={column name={$\chi_B$}, column type/.add={}{|}, /pgf/number format/fixed zerofill=false},%
        columns/xiSL/.style={column name={$\xi_\text{SL}$}},%
        columns/ESL/.style={column name={$E_\text{SL}$}},%
        columns/mSL/.style={column name={$m_\text{SL}$}},%
        columns/xiTL/.style={column name={$\xi_\text{TL}$}},%
        columns/ETL/.style={column name={$E_\text{TL}$}},%
        columns/mTL/.style={column name={$m_\text{TL}$}},%
        clear infinite=true,%
    ]{data/data_TL.txt}
    \caption{Ground state energies $E_0$ and magnetic order parameter $m$ for the triangular lattice Heisenberg antiferromagnet obtained by variational optimization on the SL and TL. While the correlation lengths have been evaluated in the limit $\chi_E \rightarrow \infty$, expectation values are computed for large, but finite $\chi_E$ where convergence is reached.}
    \label{tab:energyOrderParameter_TL}
\end{table}
These values can finally be extrapolated to the limit of infinite PEPS bond dimension $\chi_B$ using the infinite-$\chi_E$ limit of the correlation length~\cite{Corboz2018,Vanhecke2022}, as shown in Fig.~\ref{fig:results_TL}. 
Our SL CTMRG results successfully reproduce the ground state energy of $E = \pgfmathprintnumber[/pgf/number format/fixed,/pgf/number format/precision=5,/pgf/number format/fixed zerofill=true]{\zeroTLSL}$ and the magnetic order parameter of $m = \pgfmathprintnumber[/pgf/number format/fixed,/pgf/number format/precision=3,/pgf/number format/fixed zerofill=true]{\zeroMSL}$, consistent with previous infinite PEPS studies~\cite{Li2022,Hasik2024}. 
Our new TL CTMRG approach yields substantially improved variational results at the same bond dimension, which nicely align with the SL results. 
In particular, the energy $E = \pgfplotstablegetelem{3}{ETL}\of{\datatableTL}\pgfmathprintnumber[/pgf/number format/fixed,/pgf/number format/precision=5,/pgf/number format/fixed zerofill=true]{\pgfplotsretval}$ at $\chi_B = 5$ represents the lowest variational tensor network value reported to date, improving upon both extrapolated DMRG~\cite{Huang2024} and previous variational PEPS calculations~\cite{Hasik2024}. 
Although the TL PEPS ansatz has a better expressivity already at small bond dimensions, we need to neglect the lowest $\chi_B = 2$ data point in the final extrapolations. 
Our estimate for the ground state energy of the triangular lattice Heisenberg model in the infinite bond dimension limit is $E = \pgfmathprintnumber[/pgf/number format/fixed,/pgf/number format/precision=5,/pgf/number format/fixed zerofill=true]{\zeroTLTL}$, in good agreement with the SL result. 
The magnetization however shows a systematically lower value compared to the SL case, resulting in an extrapolated value of $m = \pgfmathprintnumber[/pgf/number format/fixed,/pgf/number format/precision=3,/pgf/number format/fixed zerofill=true]{\zeroMTL}$. 
This can be attributed to the improved entanglement structure, to which the order parameter seems more sensitive than the energy. 

Overall, the TL PEPS results provide a slight improvement over the SL based simulations, and are generally in good agreement with previous tensor network studies in the literature.

\subsection{Kagome lattice Heisenberg model}

\begin{figure*}[ht]
    \centering
    \tikzsetnextfilename{results_KL}
\begin{tikzpicture}
    
    \begin{axis}[
        xlabel=$1/\xi^3$,
        xlabel shift=-2,
        ylabel=$E_0$,
        ylabel shift=-2,
        width=0.99\columnwidth,
        height=0.8\columnwidth,
        legend style={font=\footnotesize},
        label style={font=\small},
        tick label style={font=\footnotesize},
        every axis/.append style={thick},
        /tikz/mark size=1pt,
        legend pos=north west,
        xmajorgrids=true,
        ymajorgrids=true,
        xmin = 0.0, 
        xmax = 1.5, 
        xtick = {0.0, 0.3, 0.6, 0.9, 1.2, 1.5}, 
        xticklabels = {$0.0$, $0.3$, $0.6$, $0.9$, $1.2$, $1.5$}, 
        ymin = -0.442,
        ymax = -0.40,
        ytick = {-0.44, -0.43, -0.42, -0.41, -0.40},
        yticklabels = {$-0.44$, $-0.43$, $-0.42$, $-0.41$, $-0.40$}, 
    ]

    \addplot[
        color = mathmColor1, 
        mark = square, 
        mark options=solid,
        line width=1.5pt, 
        mark size=1.5pt,
        only marks,
        point meta = explicit symbolic,
        visualization depends on = {value \thisrow{D} \as \mylabel},
        visualization depends on={\thisrow{angleSL} \as \myangle},
        every node near coord/.append style={%
            anchor=north east, label={\myangle:{\mylabel}}%
        },
        nodes near coords,
        legend image post style={sharp plot}
    ]
    table[
        col sep=&,%
        x expr = 1/((2*\thisrow{xiSL})^3),
        y = ESL
    ] {data/data_KL.txt};

    \addlegendentry{SL PEPS}

    \addplot[
        color = mathmColor2, 
        mark = triangle, 
        line width=1.5pt, 
        mark size=1.5pt,
        only marks,
        point meta = explicit symbolic,
        visualization depends on = {value \thisrow{D} \as \mylabel},
        visualization depends on={\thisrow{angleTL} \as \myangle},
        every node near coord/.append style={%
            anchor=north east, label={\myangle:{\mylabel}}%
        },
        nodes near coords,
        legend image post style={sharp plot}
    ]
    table[
        col sep=&,%
        x expr = 1/((2*\thisrow{xiTL})^3),
        y = ETL
    ] {data/data_KL.txt};
    \addplot[
        draw = none,
        no markers,
        forget plot
    ]
    table[
        col sep=&,%
        x expr = 1/((2*\thisrow{xiTL})^3),
        y = {create col/linear regression = {y = ETL}}
    ] {data/data_KL.txt};
    \xdef\zeroKLTL{\pgfplotstableregressionb};
    
    \addplot[mathmColor2, domain=0.0 : 1.5, forget plot] {\pgfplotstableregressiona * x + \zeroKLTL};
    \addlegendentry{TL PEPS ($\rightarrow \pgfmathprintnumber[/pgf/number format/fixed,/pgf/number format/precision=5,/pgf/number format/fixed zerofill=true]{\zeroKLTL}$)};

    \addplot[mark=none, mathmColor3, domain=0:1.5] {-0.43752};
    \addlegendentry{iPESS~\cite{Liao2017}};

    \addplot[mark=none, mathmColor4, domain=0:1.5] {-0.4386};
    \addlegendentry{DMRG~\cite{Depenbrock2012}};

    \addplot[mark=none, mathmColor5, domain=0:1.5] {-0.438703897};
    \addlegendentry{ED~\cite{Laeuchli2019}};

    \end{axis}

    \begin{axis}[
        xlabel=$1/\xi$,
        xlabel shift=-2,
        ylabel=$m$,
        ylabel shift=-2,
        width=0.99\columnwidth,
        height=0.8\columnwidth,
        legend style={font=\footnotesize},
        label style={font=\small},
        tick label style={font=\footnotesize},
        every axis/.append style={thick},
        /tikz/mark size=1pt,
        legend pos=north west,
        xmajorgrids=true,
        ymajorgrids=true,
        xmin = 0.0, 
        xmax = 1.2, 
        xtick = {0.0, 0.3, 0.6, 0.9, 1.2}, 
        xticklabels = {$0.0$, $0.3$, $0.6$, $0.9$, $1.2$}, 
        ymin = 0.00, 
        ymax = 0.32, 
        ytick = {0.00, 0.05, 0.10, 0.15, 0.20, 0.25, 0.30},
        yticklabels = {$0.00$, $0.05$, $0.10$, $0.15$, $0.20$, $0.25$, $0.30$},
        xshift=0.99\columnwidth,
    ]

    \addplot[
        color = mathmColor1, 
        mark = square, 
        line width=1.5pt, 
        mark size=1.5pt,
        only marks,
        point meta = explicit symbolic,
        visualization depends on = {value \thisrow{D} \as \mylabel},
        visualization depends on={\thisrow{angleSL} \as \myangle},
        every node near coord/.append style={%
            anchor=north east, label={\myangle:{\mylabel}}%
        },
        nodes near coords
    ]
    table[
        col sep=&,%
        x expr = 1/(2*\thisrow{xiSL}),
        y expr = 1*(\thisrow{mSL})
    ] {data/data_KL.txt};
    \addlegendentry{SL PEPS};



    \addplot[
        color = mathmColor2, 
        mark = triangle, 
        line width=1.5pt, 
        mark size=1.5pt,
        only marks,
        point meta = explicit symbolic,
        visualization depends on = {value \thisrow{D} \as \mylabel},
        visualization depends on={\thisrow{angleTL} \as \myangle},
        every node near coord/.append style={%
            anchor=north east, label={\myangle:{\mylabel}}%
        },
        nodes near coords
    ]
    table[
        col sep=&,%
        x expr = 1/(2*\thisrow{xiTL}),
        y expr = 1*(\thisrow{mTL})
    ] {data/data_KL.txt};
    \addplot[
        draw = none,
        no markers,
        forget plot
    ]
    table[
        col sep=&,%
        x expr = 1/((2*\thisrow{xiTL})),
        y = {create col/linear regression = {y = mTL}}
    ] {data/data_KL.txt};

    \addplot[mathmColor2, dotted, domain=0.0:1.2, forget plot] {\pgfplotstableregressiona * x + \pgfplotstableregressionb};
    \addlegendentry{TL PEPS};


    \end{axis}

\end{tikzpicture}
    \caption{Ground state energy $E_0$ and magnetic order parameter $m$ for the nearest-neighbour Heisenberg model on the kagome lattice. The data points are labeled by the PEPS bond dimension $\chi_B$ for a better comparison. The energy is compared against iPESS~\cite{Liao2017}, DMRG~\cite{Depenbrock2012} and exact diagonalization~\cite{Laeuchli2019}.}
    \label{fig:results_KL}
\end{figure*}
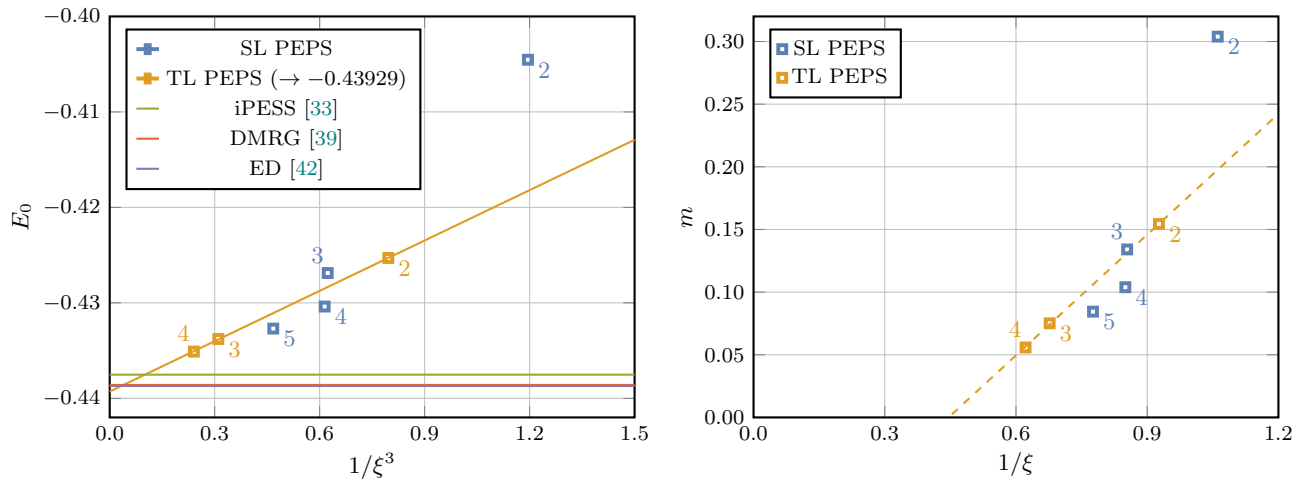

Now let us turn to the kagome Heisenberg antiferromagnet, one of the most celebrated examples of a highly frustrated quantum magnet and numerical benchmark. 
In contrast to the triangular lattice, the kagome system does not exhibit magnetic long-range order down to zero temperature and is believed to realize a QSL ground state. 
Upon coarse-graining of the three-site basis (cf. Fig.~\ref{fig:kagomeLattice_9}), the model can be variationally optimized directly on its underlying triangular lattice with a single-site PEPS ansatz. 
Similar to our findings for the triangular lattice Heisenberg model, the optimization on the native TL provides a systematic improvement over the conventional SL formulation. 
As shown in Table~\ref{tab:energyOrderparameter_KL}, the TL optimization consistently yields lower variational energies across all accessible bond dimensions $\chi_B$. 

\begin{table}[ht]
    \centering
    \begin{filecontents*}{data/data_KL.txt}
D & xiSL & ESL & mSL & xiTL & ETL & mTL & angleSL & angleTL
2 & 0.47112617716923777 & -0.40453482 & 0.30388082 & 0.539457216759 & -0.42529974 & 1.54353085e-01 & 0 & 0
3 & 0.5853781235094372 & -0.42687713 & 1.34066358e-01 & 0.7384737403406677 & -0.43378523 & 7.51480223e-02 & 90 & 0
4 & 0.5881788171816832 & -0.43038424 & 1.03910122e-01 & 0.8039310486778114 & -0.43510411 & 5.58570726e-02 & 0 & 90
5 & 0.6442477334052328 & -0.43269346 & 8.44388255e-02 & nan & nan & nan & 0 & 0
\end{filecontents*}
\pgfplotstabletypeset[%
        col sep=&,%
        columns={D, xiSL, ESL, mSL, xiTL, ETL, mTL},%
        /pgf/number format/fixed,%
        /pgf/number format/precision=5,%
        /pgf/number format/fixed zerofill=true,%
        every head row/.style={after row=\hline},%
        columns/D/.style={column name={$\chi_B$}, column type/.add={}{|}, /pgf/number format/fixed zerofill=false},%
        columns/xiSL/.style={column name={$\xi_\text{SL}$}},%
        columns/ESL/.style={column name={$E_\text{SL}$}},%
        columns/mSL/.style={column name={$m_\text{SL}$}},%
        columns/xiTL/.style={column name={$\xi_\text{TL}$}},%
        columns/ETL/.style={column name={$E_\text{TL}$}},%
        columns/mTL/.style={column name={$m_\text{TL}$}},%
        clear infinite=true,%
    ]{data/data_KL.txt}
    \caption{Ground state energies $E_0$ and magnetic order parameter for the kagome lattice Heisenberg antiferromagnet obtained by variational optimization on the SL and TL. While the correlation lengths have been evaluated in the limit $\chi_E \rightarrow \infty$, expectation values are computed for large, but finite $\chi_E$ where convergence is reached. The TL optimization for $\chi_B = 5$ is currently inaccessible due to the exceedingly large environment bond dimension required.}
    \label{tab:energyOrderparameter_KL}
\end{table}
In contrast to the triangular lattice Heisenberg model, where the nature of the ground state is well known, the situation on the kagome lattice is less clear. 
This also applies to the extrapolation of the finite bond dimension data. 
However, we find that the optimized data on the TL again allows for a reliable extrapolation in the correlation length, even for the low bulk bond dimension data accessible, cf.~Fig.~\ref{fig:results_KL}. 
In contrast, this extrapolation does not seem to be accurate for the SL data, where higher bond dimensions are required. 
This observation can be traced back to the improved representation of the ground state and its entanglement structure for the TL PEPS ansatz, and possibly the additional coarse-graining. 

Our ground state energy extrapolation yields \mbox{$E = \pgfmathprintnumber[/pgf/number format/fixed,/pgf/number format/precision=5,/pgf/number format/fixed zerofill=true]{\zeroKLTL}$}, slightly improving on previous DMRG~\cite{Depenbrock2012} and ED~\cite{Laeuchli2019} results. 
While both optimization schemes support a non-magnetic ground state at large bulk bond dimensions consistent with a quantum spin liquid, the TL data again shows a much cleaner behaviour with a hypothetical $1/\xi$ scaling. 
We note that for the scenario of a gapped ground state, the scaling to $\xi \to \infty$ could be misleading as a finite correlation length is expected in this case. 
Nevertheless, the optimized states on the TL show a clean scaling (up to the maximal accessible bond dimension), so that the extrapolation is useful to compare our result to other methods.
The optimization on the TL hence provides a significant improvement over the mapped SL ansatz, even more than on the regular triangular lattice. 
This could be attributed to the different nature of the ground state, or impacted by the additional coarse-graining. 
Since there is no clear consensus about the gapped or gapless nature of the ground state yet, it would be interesting to probe low-energy excitations~\cite{Ponsioen2022,Ponsioen2023,Tu2024} using PEPS on the triangular lattice. 

\section{Conclusion and outlook}

In this work, we have introduced a general and versatile corner transfer matrix renormalization group algorithm tailored to infinite projected entangled-pair states on the triangular lattice. 
By suitably incorporating automatic differentiation techniques for variational energy optimization, we can directly target ground states on the native lattice, eliminating the need for indirect mappings to square lattice structures. 
This approach significantly improves variational accuracy at fixed bond dimensions, as demonstrated for the triangular and kagome lattice antiferromagnetic Heisenberg models. 
These improvements can be attributed to the increased number of variational parameters and the more accurate representation of the entanglement structure in all directions, boosting the expressiveness of the ansatz. 
Although the larger coordination number increases the computational cost, we demonstrate that meaningful and competitive simulations remain feasible. 
The presented results highlight the potential of our method to push beyond previous variational benchmarks and to provide an unbiased perspective on open problems in frustrated quantum magnetism.

Beyond benchmarking, the presented method opens up a variety of promising directions. 
In terms of the algorithm, further improvements could be achieved through more efficient environment contractions, such as the recently proposed split-CTMRG scheme~\cite{Naumann2025}, or by systematically exploiting point-group and global symmetries. 
Physically, the ability to treat next-to-nearest neighbour interactions without complex routing enables more accurate studies of other spin liquid candidates, such as the triangular $J_1–J_2$ Heisenberg model~\cite{Zhu2015,Iqbal2016}. 
Furthermore, the method lends itself well to other triangular-based lattices with only a moderate coarse-graining overhead, including the honeycomb or trellis (two-site basis), dice (three-site basis) and maple-leaf lattice (six-site basis). 
The triangular CTMRG algorithm is a powerful and versatile tool for simulating strongly correlated quantum systems on frustrated lattices, paving the way for future explorations of exotic quantum phases and challenging numerical regimes.

\paragraph*{Code availability.}
A variational PEPS implementation on the triangular lattice is available in our open source variPEPS library~\cite{variPEPS_GitHub, naumann24_varipeps_python}. 
The data that support the findings of this article are openly available~\cite{naumann25_data_tl_ctmrg}.


\begin{acknowledgments}
We would like to thank Juraj Hasik, Roberto Losada, Matteo Rizzi and Erik L.\ Weerda for inspiring discussions. 
This work has been funded by the Deutsche Forschungsgemeinschaft (DFG, German Research Foundation) under the project number 277101999 -- CRC 183 (projects A04 and B01), the BMFTR (MUNIQC-Atoms, FermiQP), the Munich Quantum Valley and Berlin Quantum. 
This work has been supported by Horizon Europe programme HORIZON-CL4-2022-QUANTUM-02-SGA via the project 101113690 (PASQuanS2.1). It has also received funding from the Clusters of Excellence MATH+ and ML4Q.
\end{acknowledgments}

\bibliography{references}

\newpage
\appendix*

\section{CTMRG details}
\label{appendix}

In this appendix, we describe in detail the full triangular CTMRG procedure. 
This includes the absorption steps for both corner and edge tensors, as well as the calculation of projectors. 
Contrary to the square lattice CTMRG, where the absorption steps can be done \emph{directionally} due to the orthogonal lattice directions, the absorptions in the triangular lattice CTMRG need to be done \emph{radially}, as we describe below.

\subsection{Absorption steps}

In this section, we will describe a radial absorption step of local PEPS tensors into the environment. 
Naturally, this comes at the expense of increasing the environment bond dimension, which has to be truncated back to a fixed value $\chi_E$ to avoid an exponential increase of the computational cost and hence keep the simulations manageable. 
As in the regular square lattice CTMRG we make use of projectors for this renormalization task, whose calculation is described in the next section. 
However, the absorptions in the triangular lattice CTMRG are slightly more involved compared to the square lattice case, as updates of the corner tensors can no longer be made independently in all six lattice directions.

Each absorption step will be labeled by a certain angle $\theta$, in which the network is grown and subsequently truncated. 
Hence, absorptions into corner tensors correspond to angles $\theta_C = 2n \cdot 30^\circ$ while absorptions into edge tensors correspond to angles $\theta_T = (2n+1) \cdot 30^\circ$, for $n \in \lbrack 1, 6 \rbrack$. 
In the following, we omit the degree \mbox{symbol $(^\circ)$} and label all directional moves and corresponding tensor network diagrams simply by the integer value of the angle.

\begin{figure}[b!]
    \centering
    \includegraphics[scale = 1.0]{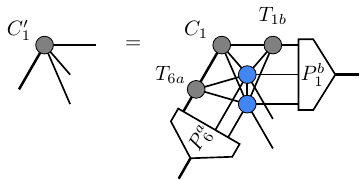}
    \caption{Absorption of the PEPS tensor, its conjugate and two edge tensors into the corner tensor $C_1$. The enlarged spaces are renormalized to the environment bond dimension $\chi_E$ using left and right projectors.}
    \label{fig:absorption_C1_1}
\end{figure}

We describe the absorption and subsequent renormalization steps for corner tensor $C_1$ and edge tensors $T_{1a}$, $T_{1b}$. 
The projectors required to renormalize, i.e., truncate the enlarged spaces are assumed to be given for now, while the method to compute them is explained in detail in Sec.~\ref{sub:projectorCalculation}. 
The absorption of corner tensor $C_1$ is shown in Fig.~\ref{fig:absorption_C1_1}.
Here, the projectors belong to different directions $\theta$. 
We highlight the difference to the square lattice CTMRG, where the corner updates do not involve the absorption of a PEPS tensor.

\begin{figure}[ht]
    \centering
    \includegraphics[scale = 1.0]{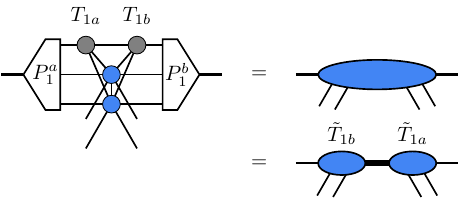}
    \caption{Absorption of the PEPS tensor and its conjugate into the edge tensors $T_{1a}$ and $T_{1b}$ and renormalization with left and right projectors. The resulting tensor is decomposed with an exact SVD into two new edge tensors. The shared index of dimension $(\chi_E \cdot \chi_B^2)$ needs to be renormalized with a second set of projectors. Note, that the labels $a$ and $b$ get exchanged due to the absorption of the PEPS tensor, see main text for details.}
    \label{fig:absorption_T1_1}
\end{figure}

\begin{figure}[ht]
    \centering
    \includegraphics[scale = 1.0]{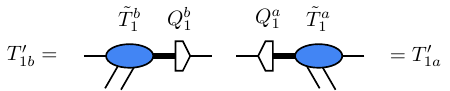}
    \caption{Renormalization of the enlarged bond between $\tilde T_{1b}$ and $\tilde T_{1a}$ by application of a second set of projectors $Q$, resulting in the updated edge tensors.}
    \label{fig:absorption_T1_2}
\end{figure}

We now proceed with the update of tensors $T_1$. 
This absorption step is different to the square lattice one as well, as both edge tensors are updated simultaneously. 
Therefore, they have to be restored from a combined six-index tensor after the absorption, which moreover needs to be done in two successive steps. 
The absorption step and the first decomposition into tensors $(\tilde T_{1a}, \tilde T_{1b})$ is shown in Fig.~\ref{fig:absorption_T1_1}. 
Here, the combined tensor is decomposed using an SVD, where the singular values are evenly split to each $\tilde T$ tensor. 
Notice, that the order of the edge tensors flips, i.e.,  $a \leftrightarrow b$, due to the absorption of the PEPS layer. 
Unfortunately, in this step we cannot simply reduce the enlarged dimension of size $(\chi_E \cdot \chi_B^2)$ back to just $\chi_E$, as this truncation of the singular values induces a significant error and leads to an unstable CTMRG routine, as also noted in Refs.~\cite{Lukin2024,Naumann2025}. 
For this reason, we need to compute a second set of truncation projectors $Q$ to be inserted into the connecting bond. 
These projectors must be computed from a larger patch of the infinite tensor network in order to truncate to the relevant subspace. 
A direct SVD, as mentioned above, is too local to perform this task accurately. 
However, naively truncated tensors $\tilde T_{1}^\text{trunc}$ can still be used in order to compute accurate and stable projectors $Q$, as described in Sec.~\ref{sub:projectorCalculation}. 
Ultimately, with the second set of projectors computed, the updated edge tensors $T_{1a}$ and $T_{1b}$ are obtained by a simple contraction as shown in Fig.~\ref{fig:absorption_T1_2}.
Compared to the square lattice CTMRG variant, the absorptions in different directions are no longer independent from each other due to the non-orthogonal lattice vectors. 
Therefore, a full CTMRG iteration performs a radial absorption scheme, as outlined below.
\begin{enumerate}
    \item[1.] Compute all pairs of projectors $(P_{i}^a, P_{i}^b)$.
    \item[2.] Perform all corner absorptions and renormalizations to obtain new $C_{i}^\prime$, cf.\ Fig.~\ref{fig:absorption_C1_1}.
    \item[3.] Perform all edge absorptions to obtain $(\tilde T_{ia}, \tilde T_{ib})$ and their preliminary truncated versions, cf.\ Fig.~\ref{fig:absorption_T1_1}.
    \item[4.] Compute all pairs of projectors $(Q_{i}^a, Q_{i}^b)$.
    \item[5.] Perform renormalization of all edges to obtain $(T_{ia}^\prime, T_{ib}^\prime)$, cf.\ Fig.~\ref{fig:absorption_T1_2}. 
\end{enumerate}
The iterative CTMRG procedure is terminated once all the environment tensor converge element-wise to a preset tolerance. 

\subsection{Projector calculation}
\label{sub:projectorCalculation}

The projectors used in the previous section to renormalize the environment tensors after the absorption need to be appropriately computed in order to arrive at a stable and accurate CTMRG power method. 
To this end, we define \emph{two-third} and \emph{full projectors}, which take into account two-thirds or the full environment in all lattice directions, respectively. 
Those projectors are the counterparts of well-established projectors in the numerical PEPS community~\cite{Naumann2024}, adapted to the triangular structure.
The initial networks from which the projectors are computed consist of PEPS tensors with their respective environments, each containing a $120^\circ$ patch of the full infinite lattice. 
\begin{figure}[ht]
    \centering
    \begin{minipage}{0.9\columnwidth}
        \centering
        \includegraphics[scale = 1.0]{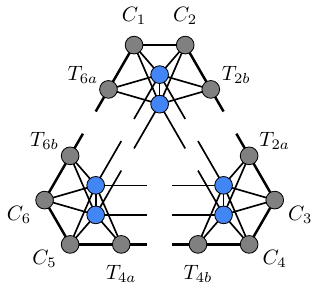}
    \end{minipage}
    \begin{minipage}{0.9\columnwidth}
        \centering
        \includegraphics[scale = 1.0]{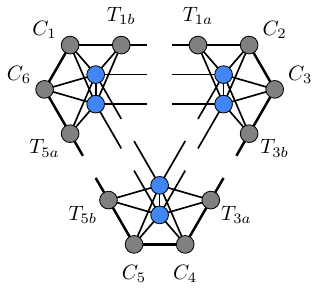}
    \end{minipage}
    \caption{The six independent patches serving as the starting point for computing truncation projectors required in the absorption process. The individual corners are contracted into a matrix $\rho_{\theta}$, where $\theta$ denotes the angle pointing towards the patch. See the main text for more details.}
    \label{fig:tensorDefinition_CTMRG_2_3}
\end{figure}
In total there are six individual patches, as shown in Fig.~\ref{fig:tensorDefinition_CTMRG_2_3}, each of which can be contracted to form a matrix $\rho_{\theta}$ of dimension $(\chi_E \cdot \chi_B^2) \times (\chi_E \cdot \chi_B^2)$. 
Hence, we define six matrices $\rho_{\theta}$, with $\theta \in \lbrack 30, 90, 150, 210, 270, 330 \rbrack$. 
The six different sets of projectors required to update all possible environment tensors in the CTMRG fixed point routine are computed for the links connecting $T_{ib}$ and $T_{ia}$, see also Fig.~\ref{fig:tensorDefinition_CTMRG_0}. 
Hence, the direction for which to compute projectors are for the same values of angles $\theta$. 
We exemplify the projector calculation for the renormalization step of edge tensors $T_{1}$, as shown in Fig.~\ref{fig:absorption_T1_1}. 
However, most definitions are valid for any absorption direction.

\begin{figure}[ht]
    \centering
    \includegraphics[scale = 1.0]{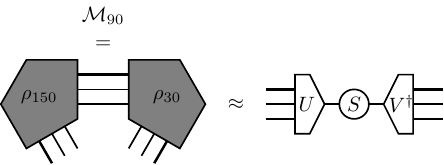}
    \caption{Definition of the matrix $\mathcal M_{90}$ used to compute two-third projectors at angle $\theta = 90$. The matrix is decomposed using a truncated singular value decomposition.}
    \label{fig:projectorDefinition_T1_1}
\end{figure}

\begin{figure}[ht]
    \centering
    \includegraphics[scale = 1.0]{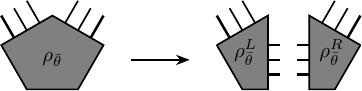}
    \caption{Decomposition of $\rho_{\bar\theta}$, here for the example of \mbox{$\theta = 90$} for the calculation of full projectors. The equal splitting makes the final projectors more balanced, although the cut is performed at an artificial position that is not an actual connection in the triangular lattice PEPS.}
    \label{fig:projectorDefinition_T1_4}
\end{figure}

In order to compute two-third projectors for any direction $\theta$, we start by multiplying two of the $120^\circ$ patches shown in Fig.~\ref{fig:tensorDefinition_CTMRG_2_3} into a matrix
\begin{align}
    \mathcal M_{\theta} \coloneqq \rho_L \cdot \rho_R = \rho_{\theta + 60} \cdot \rho_{\theta - 60},
    \label{eq:twoThirdProjects}
\end{align}
which now contains two-thirds of the infinite lattice.
This is visualized in Fig.~\ref{fig:projectorDefinition_T1_1} for a fixed angle $\theta = 90$.
To compute full projectors, we extend the network for $\mathcal M_{\theta}$ above with the third $120^\circ$ corner $\rho_{\bar\theta}$ with $\bar\theta \coloneqq\theta + 180$, where all angles are taken to be modulo $2\pi$. 
This corner is however decomposed first using a full SVD, so that it can be equally ascribed to both $\rho_L$ and $\rho_R$ to compute more balanced projectors. 
This is visualized in Fig.~\ref{fig:projectorDefinition_T1_4}.
The open cut of the resulting network is now exactly opposite of $\theta$, i.e.,  at $\bar\theta = \theta + 180$, although this is not an actual connection in the triangular lattice PEPS. 
However, this approach helps to stabilize the procedure. 
The initial network to compute full projectors is then given by
\begin{align}
    \mathcal M_\theta \coloneqq \rho_L \cdot \rho_R = \left( \rho_{\bar\theta}^L \cdot \rho_{\theta + 60} \right) \cdot \left( \rho_{\theta - 60} \cdot \rho_{\bar\theta}^R \right).
    \label{eq:fullProjects}
\end{align}
In order to find the best possible set of projectors $P_{1}^a$ and $P_{1}^b$, we demand that the matrix
\begin{align}
    \mathcal M_{\theta}^P = \rho_L \cdot P_{1}^b \cdot P_{1}^a \cdot \rho_R \approx \mathcal M_\theta
\end{align}
with inserted projectors is an optimal approximation to the original matrix, that minimizes the Frobenius norm $\vert\vert \mathcal M_{\theta}^P - \mathcal M_{\theta} \vert\vert_F$. 
To this end, we compute the \emph{singular value decomposition} (SVD) of $\mathcal M_{\theta}$, i.e.,
\begin{align}
    \mathcal M_{\theta} = U \cdot S \cdot V^\dagger,
\end{align}
as shown in Fig.~\ref{fig:projectorDefinition_T1_1}. 
The inverse of this matrix can be inserted back into the tensor network we started 
from, which leads to an identity condition for the bonds which we aim to renormalize. 
Specifically, we have
\begin{align}
    \begin{split}
        \mathbb I &\simeq \rho_R \cdot \mathcal M_{\theta}^{-1} \cdot \rho_L \\ 
        &= \left( \rho_R V S^{-1/2} \right) \cdot \left( S^{-1/2} U^\dagger \rho_L \right) \\
        &= P_{1}^{b} \cdot P_{1}^{a},
    \end{split}
    \label{eq:definitionProjectors_1}
\end{align}
from which we can define the final projectors. 
Here $P_{1}^{a}$ acts on the left indices of $T_{1a}$, while $P_{1b}$ acts on the right indices of $T_{1b}$, as seen in Fig.~\ref{fig:absorption_T1_1}. 
Pictorially, the projectors are given by
\begin{align}
    \begin{split}
        \includegraphics[scale = 1.0]{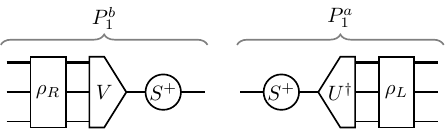}
    \end{split},
\end{align}
where $S^+ = \text{inv}(\sqrt{S})$. 
The exact identity in Eq.~\eqref{eq:definitionProjectors_1} only holds for a full SVD without any truncation. 
In practice though, the number of singular values is truncated to at most $\chi_E$, so that the projectors implement the desired renormalization step from an increased dimension \mbox{$(\chi_E \cdot \chi_B^2)$} back to $\chi_E$.

\begin{figure}[thb]
    \centering
    \includegraphics[scale = 1.0]{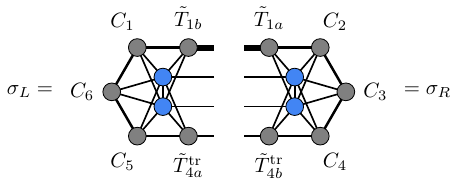}
    \caption{Definition of the networks $\sigma_L$ and $\sigma_R$, from which we compute projectors $Q_{1}^{a}$ and $Q_{1}^{b}$ to be inserted into the enlarged bond.}
    \label{fig:projectorDefinition_T1_2}
\end{figure}
Finally, we need to compute the second set of projectors $Q$, that renormalize the enlarged bonds between the partially updated edge tensors $\tilde T$. 
To this end we start with a two-site network, choosing PEPS tensors in the lattice direction parallel to the bonds between $\tilde T_{ia}$ and $\tilde T_{ib}$ that we wish to truncate. 
Proceeding with the above example of tensors $T_{1}$, the network from which to compute projectors is given in Fig.~\ref{fig:projectorDefinition_T1_2}. 
The best accuracy in this step could be achieved by also using the updated tensors $(\tilde T_{4a}, \tilde T_{4b})$ at the opposite edge. 
However, since they have an index of dimension $(\chi_E \cdot \chi_B^2)$ too, the computational cost of the successive SVD would be too high for values \mbox{$(\chi_B, \chi_E)$} of interest. 
Therefore, we use a truncated version of those edge tensors as shown in the figure, with only $\chi_E$ singular values kept. 
Although this is the only rather uncontrolled approximation in the networks for $\sigma_L$ and $\sigma_R$, it was found to work reliably in the numerical simulations.
Following the same spirit as for the first set of projectors $P$, we now contract the two networks over the enlarged index and decompose the resulting matrix according to
\begin{align}
    \Sigma = \sigma_L \cdot \sigma_R \approx U \cdot S \cdot V^\dagger,
\end{align}
where the singular values are truncated to the final bond dimension $\chi_E$. 
Inserting the inverse expression back into the network, we can define projectors according to
\begin{align}
    \begin{split}
        \mathbb I &\simeq \sigma_R \cdot \Sigma^{-1} \cdot \sigma_L \\ 
        &= \left( \sigma_R V S^{-1/2} \right) \cdot \left( S^{-1/2} U^\dagger \sigma_L \right) \\
        &= Q_{1}^{b} \cdot Q_{1}^{a}.
    \end{split}
    \label{eq:definitionProjectors_2}
\end{align}
Using the same definition $S^+ = \text{inv}(\sqrt{S})$ as above, the projectors $Q$ are given by
\begin{align}
    \begin{split}
        \includegraphics[scale = 1.0]{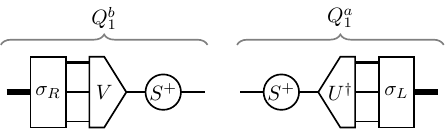}
    \end{split}.
\end{align}
They will implement the second renormalization step of the edge tensors, as shown in Fig.~\ref{fig:absorption_T1_2}.

We want to note that it is straightforward to define a potentially more stable variation of the projectors applying the proposal by Fishman et.\ al.~\cite{Fishman2018}.

\begin{figure*}[hbtp]
    \centering
    \begin{minipage}{0.32\textwidth}
        \includegraphics[scale = 1.0]{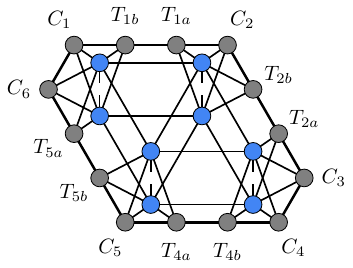}    
    \end{minipage}
    \begin{minipage}{0.32\textwidth}
        \includegraphics[scale = 1.0]{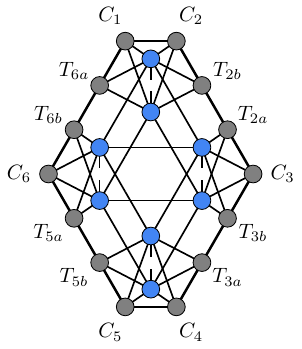}    
    \end{minipage}
    \begin{minipage}{0.32\textwidth}
        \includegraphics[scale = 1.0]{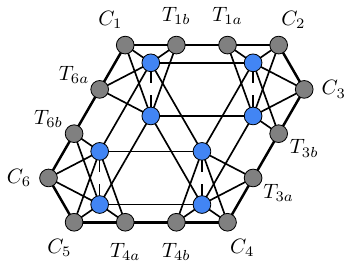}    
    \end{minipage}
    \caption{Reduced density matrices of four sites for the three different arrangements of upward and downward triangles. Those networks are used to compute next-to-nearest neighbour or plaquette expectation values on the triangular lattice.}
    \label{fig:nextNearestNeighbourTerms}
\end{figure*}

\subsection{Expectation values}

The fixed-point environment tensors can finally be used to compute expectation values, such as the ground state energy or local magnetization. 
To this end we compute $n$-site reduced density matrices $\rho_n$ from the PEPS tensors along with their respective environments. 
Expectation values are then estimated as
\begin{align}
    \langle \psi \vert O_n \vert \psi \rangle = \frac{\text{tr}(O_n \rho_n)}{\text{tr}(\rho_n)},
\end{align}
where $O_n$ is an operator acting on $n$ sites. 
A single-site reduced density matrix can be computed similarly to Fig.~\ref{fig:tensorDefinition_CTMRG_1}, just leaving open the physical indices. 
Nearest-neighbour expectation values are given by two-site expectation values in the three lattice directions $\mathbf a_1$, $\mathbf a_2$ and \mbox{$\mathbf a_1$ + $\mathbf a_2$}. 
Importantly, next-to-nearest neighbour interaction terms can be computed with networks of $2 \times 2$ PEPS and surrounding CTMRG tensors, unlike in the regular square lattice CTMRG where the mapping leads to patches of $2 \times 3$ and $3 \times 2$ tensors. 
They are depicted in Fig.~\ref{fig:nextNearestNeighbourTerms}.
This reduces the amount of required memory in building the gradient for these expectation values.

\subsection{Correlation length}

The finite environment bond dimension $\chi_E$ introduces an effective length scale in the system, the correlation length $\xi(\chi_E)$. 
It can be computed from the dominant gap in the eigenvalue spectrum of the transfer matrix $\mathcal T$. 
Following the common procedure~\cite{Rader2018}, the dominant eigenvalues $\lambda_i$ corresponding to eigenvectors $\nu_i$ of $\mathcal T$ can be computed using the triangular lattice CTMRG tensors according to
\begin{align}
    \begin{split}
        \includegraphics[scale = 1.0]{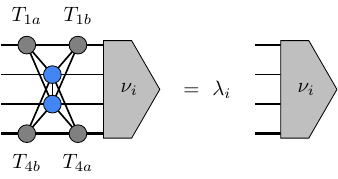}
    \end{split}.
    \label{eq:correlationLength}
\end{align}
Naturally, the row of PEPS and CTMRG tensors needs to respect the full periodicity of the unit cell. 
Since the PEPS ansatz and hence the environment tensors are not enforced to be rotation symmetric, the correlation length is computed in each direction $\mathbf a_1$, $\mathbf a_2$ and $\mathbf a_1 + \mathbf a_2$ of the triangular lattice and we use the dominant one for further analysis. 
For the eigenvalue solver, the initial boundary tensor $\nu$ can be computed from the environment itself, using $T_{1a}$, $C_2$, $C_3$, $C_4$ and $T_{4b}$, alongside the local PEPS tensor in the example of Eq.~\eqref{eq:correlationLength}.

\end{document}